\documentclass[aps,physrev,twocolumn,superscriptaddress]{revtex4-2}

\usepackage{graphicx} %allows use of postscript graphics and color output

\usepackage{amsmath}

\begin{document}

%\preprint{}

\title{High-bandwidth Coherence Cloning using Optical-Phase-Locking Feedforward}

\author{Chen Jia}
\thanks{These authors contributed equally to this work.}
\affiliation{State Key Laboratory of Low Dimensional Quantum Physics, Department of Physics, Tsinghua University, Beijing 100084, China.}

\author{Zhen-Xing Hua}
\thanks{These authors contributed equally to this work.}
\affiliation{State Key Laboratory of Low Dimensional Quantum Physics, Department of Physics, Tsinghua University, Beijing 100084, China.}

\author{Yu-Xin Chao}
\affiliation{State Key Laboratory of Low Dimensional Quantum Physics, Department of Physics, Tsinghua University, Beijing 100084, China.}
\author{Meng Khoon Tey}
\email{mengkhoon\_tey@mail.tsinghua.edu.cn}
\affiliation{State Key Laboratory of Low Dimensional Quantum Physics, Department of Physics, Tsinghua University, Beijing 100084, China.}
\affiliation{Frontier Science Center for Quantum Information, Beijing, China.}
\affiliation{Hefei National Laboratory, Hefei, Anhui 230088, China.}


\date{\today}

\begin{abstract}
Ultra-narrow-linewidth lasers with suppressed high-frequency phase noise are critical for quantum control and precision metrology. While optical phase locking (OPL) is the standard technique for cloning the coherence of such sources, its effectiveness is often limited at high frequencies by feedback latency. We present a robust feedforward architecture that overcomes this limitation by recycling and demodulating the existing master-slave beat signal to drive a single electro-optic modulator for near-instantaneous noise cancellation. This approach eliminates the extraneous sidebands and transmission losses typical of more complex modulators. Through active stabilization of the beat amplitude and demodulation phase, we demonstrate robust suppression exceeding 30 dB from 10 kHz to 10 MHz. This hardware-efficient framework is readily compatible with standard OPL setups, offering a scalable solution for high-fidelity coherent control.
\end{abstract}

%\keywords{}

\maketitle

\section{Introduction}
\label{sec:introduction}

Laser systems featuring ultra-narrow linewidths and suppressed high-frequency phase noise are foundational to various frontier applications such as optical clocks~\cite{2025_Hassan_Beloy_Siegel_Kobayashi_Swiler_Grogan_Brown_Rojo_Bothwell_Hunt_Cryogenic_Optical_Lattice_Clock}, gravitational wave detection~\cite{2015_LIGO_AdvancedLIGO,2016_Ye_GWD_with_atomic_clocks}, photonic microwave and THz wave generation~\cite{2023_Djevahirdjian_Lechevallier_Martin-Drumel_Pirali_Ducournau_Kassi_Frequency_stable_and_low_phase_noise_THz,2025_He_NatComm_microwave_generation} and precision control of ions~\cite{2021_Egan_Debroy_Noel_Risinger_Zhu_Biswas_Newman_Li_Brown_Cetina_Fault-tolerant_control,2025_Smith_Single-Qubit_Gates_with_Errors_at_10-7}, neutral atoms~\cite{2025_Ye_PRX_High_Power_Clock_Laser,2023_Evered_High-fidelity_parallel_entangling_gates,2024_Bluvstein_Logical_quantum_processor_based,2022_Zhan_Rydberg_gate}, and molecules~\cite{2014_Molony_Creation_of_Ultracold}. While applications such as optical clocks prioritize sub-hertz linewidths for long-duration interrogation~\cite{2017_YeJun_mHzlaser,2025_Xibo_narrowlinewidthcavity}, fast quantum gates necessitate both frequency stability and minimal phase deviations within the gate duration to suppress operational errors~\cite{2025_Ye_PRX_High_Power_Clock_Laser}. However, developing these high-performance systems is costly and complex, typically requiring light sources with sub-tens of kHz free-running linewidths, ultra-stable high-finesse cavities  for linewidth reduction, and rigorous calibration against narrow atomic transitions. 

Given the complexity of these laser systems, it is of significant interest to seek more economical methods to replicate the temporal coherence of an existing high-quality laser at different frequencies. Optical phase-locking is an effective solution for this purpose, routinely yielding sub-Hz relative linewidths~\cite{1963_Rabinowitz_AFC_Optical_Heterodyne_Detector,1965_Enloe_Laser_phase-locked_loop,1983_Steele,1987_Hall_Principles_of_optical_phase-locking,1999_Ye_Hall_Optical_phase_locking_in_the_microradian_domain,2018_Brunel_subHz_PLL}. OPL operates by monitoring the beat signal between a reference and a slave and employing a high-speed feedback loop to correct the slave's phase drift. However, the inherent latency in feedback electronics and optical paths often causes traditional OPL to amplify, rather than suppress, high-frequency phase noise beyond  a few MHz. This limitation renders OPL insufficient for slave lasers with broad technical linewidths. Recently, feedforward control has emerged as a promising alternative solution to this problem~\cite{2009_Bagheri_Semiconductor_laser_phase_noise_cancellation, 2012_Aflatouni_Wideband_tunable_laser_phase_noise_reduction, 2016_Okamoto_Toge_Manabe_Robust_Laser_Phase_Noise_Reduction,2019_HeZuyuan_AOM, 2017_Lintz_Note_MZI_FFFB, 2025_Leseleuc_FF,2024_Chao_PDHFeedForward,2025_Chao_Robust,2025_Hua_Feedforward_FDL,2010_Koke_NatPhys_AOM_FF_Comb,2012_Sala_Wide-bandwidth_phase_lock, 2015_Burkart_Optical_phase_cloning,2016_Watts_Phase_Noise_Reduction_of_an_Optical_Frequency_Comb_beat_MZM,2017_Besnard_beat_MZM,2020-Tan-Sensor-beat-FFFB-AOM, 2022_Li_Active_Cancellation_of_Servo-Induced_Noise,2024_Beat_IQmodulator}, offering superior bandwidth and demonstrable performance gains~\cite{2025_Maddox_Enhanced_Quantum_State_Transfer}. By correcting phase errors post-source and utilizing optical delay lines to compensate for the detection and correction-signal lags, feedforward methods can, in principle, achieve near-instantaneous noise correction.

Several beat-detection-based feedforward schemes have been demonstrated in recent years, yet they face significant implementation trade-offs. Approaches utilizing acousto-optic frequency shifters (AOFSs) for phase compensation~\cite{2010_Koke_NatPhys_AOM_FF_Comb, 2012_Sala_Wide-bandwidth_phase_lock, 2020-Tan-Sensor-beat-FFFB-AOM} are constrained by a correction bandwidth below a few MHz due to slow acoustic transit times. To achieve higher speeds, GHz-class Mach-Zehnder modulators (MZMs) have been employed~\cite{2016_Watts_Phase_Noise_Reduction_of_an_Optical_Frequency_Comb_beat_MZM, 2017_Besnard_beat_MZM}; however, MZMs generate multiple sidebands that suppress noise in one sideband while amplifying it in others. To eliminate these deleterious sidebands, Burkart et al. utilized a dual-parallel MZM—the optical equivalent of a microwave IQ mixer—at the cost of increased architectural complexity and significant power loss associated with the rejected sideband~\cite{2015_Burkart_Optical_phase_cloning}.

In this work, we introduce an efficient feedforward architecture that is highly compatible with conventional OPL while significantly extending its high-frequency performance. Drawing on our previous Pound-Drever-Hall feedforward scheme~\cite{2024_Chao_PDHFeedForward}, the current method begins with a slave laser already frequency-stabilized to a reference via standard OPL. In this configuration, we repurpose the master-slave beat signal by demodulating it to extract real-time residual phase jitter. This demodulated signal, following flat-band amplification, directly drives a single electro-optic modulator (EOM) to suppress the remaining high-frequency noise. Our approach obviates the need for precise MZM bias control, avoids extraneous sideband generation, and eliminates the associated transmission losses. By stabilizing critical system parameters, we overcome the drift typically associated with feedforward systems and demonstrate robust phase noise suppression exceeding 30 dB from 10 kHz to 10 MHz.

This article is organized as follows. Section~\ref{sec:principle} describes the working principle of OPL feedforward. Section~\ref{sec:core_setup} presents the experimental setup. Section~\ref{sec:procedure} details the optimization procedures. Section~\ref{sec:performance} evaluates the performance, reliability, and tunability of our method. A comparison of different feedforward schemes is presented in Sec.~\ref{sec:comparisons}. Finally, Section~\ref{sec:summary} provides a summary.

\section{Working Principle}
\label{sec:principle}
Our feedforward approach targets residual high-frequency phase deviations in an optically phase-locked slave-master pair offset by $\Delta\omega$. Under locked conditions, the laser fields can be expressed as
\begin{equation}\label{electric_field}
 \begin{aligned}
     E_{\rm ref}=&E_{1}\sin[\omega_0
t+\phi^n_{\rm ref}(t)], \\
     E_{\rm sla}=& E_{2}\sin[(\omega_0+\Delta \omega)t+\phi^n_{\rm ref}(t)+\phi_{r}(t)].
\end{aligned}
\end{equation}
Here, $\omega_0$ and $\phi^n_{\rm ref}(t)$ denote the reference laser's center frequency and phase jitter, while $\phi_r(t)$ represents the residual high-frequency phase deviation not removed by feedback. To extract $\phi_r(t)$ for feedforward suppression, we recycle the beat signal used for OPL:
\begin{equation}\label{eq:beatSignal}
V_{\rm beat}(t)=A_{\rm beat}\cos[\Delta \omega t+\phi_r(t)],
\end{equation}
where $A_{\rm beat}=\frac{1}{2}G_{\rm PD}E_{1}E_{2}$ ($G_{\rm PD}$ is the gain in photo-detection). Demodulating $V_{\rm beat}$ with a local oscillator (LO) $\propto \sin(\Delta\omega t)$, at a suitable phase, yields a post-mixer, low-pass-filtered output
\begin{equation}\label{eq:errorSignal}
    V_{\rm error}(t)=A_{\rm error}\sin[\phi_r(t)]\approx A_{\rm error}\phi_r(t),
\end{equation}
$A_{\rm error}=\frac{1}{2}A_{\rm beat}G_{\rm mixer}$. As $V_{\rm error}(t)$ is proportional to $\phi_r(t)$ when $|\phi_r(t)|\ll1$, it can drive a fiber electro-optic modulator (with a flat voltage-to-phase response over broad frequencies) to cancel $\phi_r (t)$ in the slave laser, producing an output field
\begin{equation}\label{FF_field}
    E_{\rm out}=E_0\cos[(\omega_0+\Delta \omega) t+\phi^n_{\rm ref}(t)+\phi_r(t)- g_{\rm ff} \phi_r(t^{\prime})],
\end{equation}
where $g_{\rm ff}=G_{\rm amp}G_{\rm EOM}A_{\rm error}$ incorporates the amplifier gain $G_{\rm amp}$ and the response of the EOM,  $G_{\rm EOM}$. The prime in $t^{\prime}$ accounts for the potential timing mismatch. The perfect cancellation of $\phi_r(t)$ requires
\begin{equation}\label{eq:condition}
\left\{
 \begin{aligned}
      g_{\rm ff}&=1, \\
      t^{\prime}&=t.\\
\end{aligned}
\right.
\end{equation}
Under these conditions, the master's spectrum is duplicated in the slave but shifted by $\Delta \omega$.

It should be emphasized that the proposed method performs optimally when $|\phi_r(t)|\ll1$. Thus, the OPL feedback must suppress the low frequency phase noise sufficiently well to ensure the reliability of this method.

\section{Setup and results}

\subsection{The core setup}\label{sec:core_setup}
\begin{figure*}[!htbp]
\centering
%\hspace{13 mm }
\begin{minipage}{1.5\columnwidth}
\centering
\includegraphics[width=1.\columnwidth]{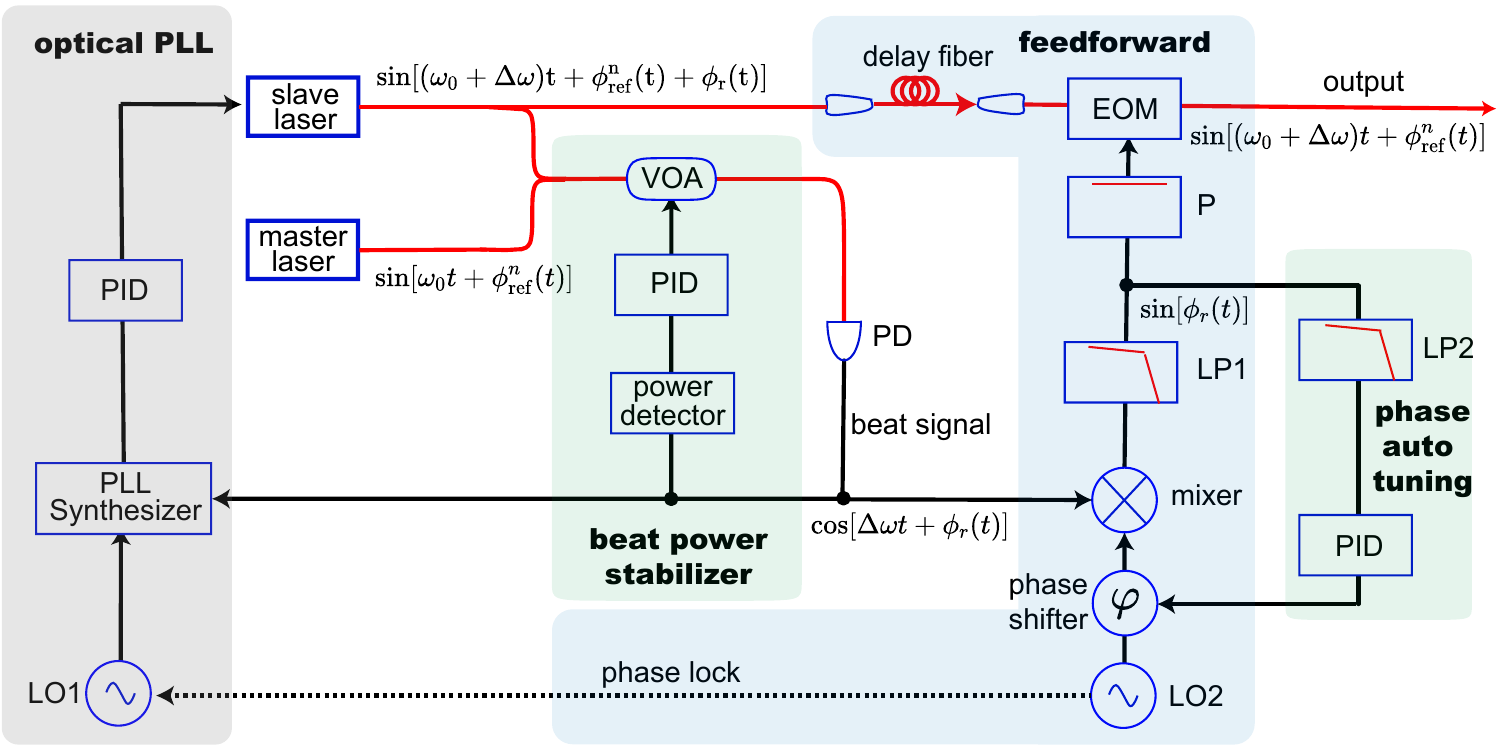}
\end{minipage}
\caption{Schematic of the OPL feedforward scheme. A standard optical phase-locked loop (gray area) locks the slave-master frequency offset, $\Delta\omega$. The feedforward stage (blue area) suppresses residual noise $\phi_r(t)$ outside the OPLL bandwidth by demodulating the photodiode (PD) beat with a second local oscillator (LO2) at frequency $\Delta \omega$, amplifying it with a constant-gain amplifier (P), and applying the resulting signal to a fiber electro-optic modulator (EOM). To ensure robustness, the beat amplitude is stabilized via a variable optical attenuator (VOA) (left green area), and the LO2 phase is adjusted to null the DC offset after the mixer (right green area). LP: low-pass filter.
}
\label{setup}
\end{figure*}

Figure~\ref{setup} illustrates the key components of our OPL feedforward setup. A photodiode (PD) detects the master-slave beat, and the resulting signal is split into three paths for the optical-phase-locked loop (OPLL), feedforward correction, and beat-amplitude stabilization:
\begin{enumerate}
\item The OPLL section utilizes a PLL synthesizer (ADF4007) to compare the phase of the beat signal against a local oscillator (LO1). A PID controller then adjusts the slave laser's current and piezoelectric transducer, yielding a feedback bandwidth of $\sim$1.8\,MHz.
\item The feedforward section comprises a mixer, a local oscillator (LO2), a low-pass filter (LP1), a flat-gain amplifier (P), and a fiber EOM. LO2 operates at $\Delta \omega$ and is phase-locked to LO1. To satisfy the quadrature condition required for extracting $\phi_r(t)$ (see Sect.~\ref{sec:principle}), we extract the DC offset from the mixer output through low-pass filtering (LP2, 8-kHz bandwidth). A PID controller then adjusts the LO2 phase to zero this offset, ensuring accurate extraction of $\phi_r(t)$ [Eq.~(\ref{eq:errorSignal})]~\cite{stabilization_note}.
\item The stabilization section maintains a constant beat amplitude. Because feedforward is an open-loop process~\cite{2025_Hua_Feedforward_FDL}, it is sensitive to drifts in $g_{\rm ff}$ caused by varying fiber transmission. We mitigate these effects by monitoring the beat strength with an RF power detector and stabilizing it via a variable optical attenuator (VOA). This approach provides a robust, cost-effective alternative to independently stabilizing the power of both lasers, assuming the beat spectrum is carrier-dominated.
\end{enumerate}

\subsection{Optimization Procedures}\label{sec:procedure}

Achieving optimal feedforward noise suppression requires precise tuning of parameters to meet the conditions in Eq.~(\ref{eq:condition}). Our procedure mirrors the general protocol outlined in Ref.~\cite{2025_Hua_Feedforward_FDL}. First, we generate and inject an artificial phase modulation at 2\,MHz into the slave beam using a signal generator (SG) and a fiber EOM (EOM1 in Fig.~\ref{fig:detailed_setup}). We then use a second SG, phase-locked to the first, to drive the feedforward fiber EOM (EOM2 in Fig.~\ref{fig:detailed_setup}) at the same frequency. We adjust the amplitude and phase of the feedforward signal until the injected phase noise is fully canceled. Once calibrated, we configure the full setup shown in Fig.~\ref{fig:detailed_setup} by adjusting the beat signal power and the gain of the feedforward amplifier P to produce the required signal amplitude. For temporal alignment, we estimate the delay in the feedforward-signal path using a vector network analyzer, select a delay fiber (7\,m in our case) slightly longer than this estimate, and fine-tune by inserting short cables between P and EOM2 until the injected noise is maximally suppressed.

\begin{figure}[htbp]
	\centering
\begin{minipage}{0.99\linewidth}
		\includegraphics[width=1\linewidth]{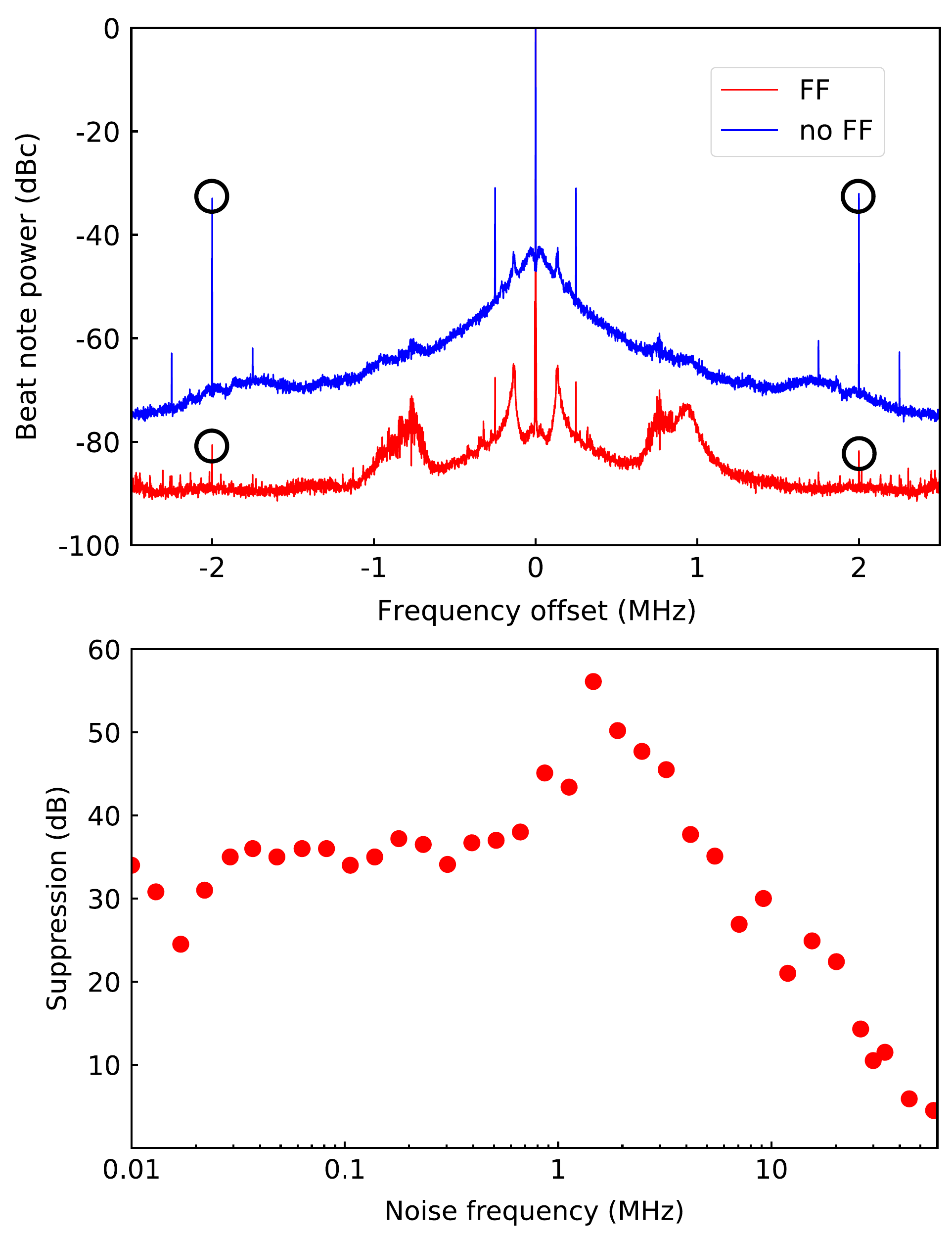}
\caption{Feedforward efficacy versus noise frequency. (a) Reference-slave beat spectra with (FF) and without (no FF) feedforward, centered at 240 MHz with a 500 Hz resolution. Injected 2 MHz sinusoidal noise powers are indicated by black circles; their ratio determines the suppression shown in (b). Peaks at $\pm$240 kHz originate from the slave laser's intrinsic modulation, which is also effectively suppressed. Residual bumps above the -90 dBc floor result from the intensity noise of the reference laser. (b) Suppression as a function of injected noise frequency.
}
\label{suppression}%
    \end{minipage}
\end{figure}

\subsection{Performance}\label{sec:performance}

We characterize the slave-master phase noise difference using a secondary beat-note setup [Fig.~\ref{fig:detailed_setup}c, blue shaded area]. To evaluate feedforward performance, we inject phase modulation into the slave via EOM1 and compare the resulting noise spectra with and without feedforward suppression for different injection frequencies. As illustrated in Fig.~\ref{suppression}, we achieve a peak suppression $>$50 dB near 1.9 MHz, with values remaining above 30 dB from 10 kHz to 10 MHz. This represents a significant improvement over our previous PDH-based method \cite{2024_Chao_PDHFeedForward}, where performance was constrained by cavity linewidth at low frequencies; for a 10 kHz linewidth cavity, suppression is typically limited to ~6 dB at 10 kHz and 20 dB at 50 kHz. The performance of our current setup is limited by the low-frequency response of the PD and mixer, and by the gain flatness of amplifier P at higher frequencies.

\begin{figure}[htbp]
	\centering
	\begin{minipage}{0.99\linewidth}
		\centering
		\includegraphics[width=1\linewidth]{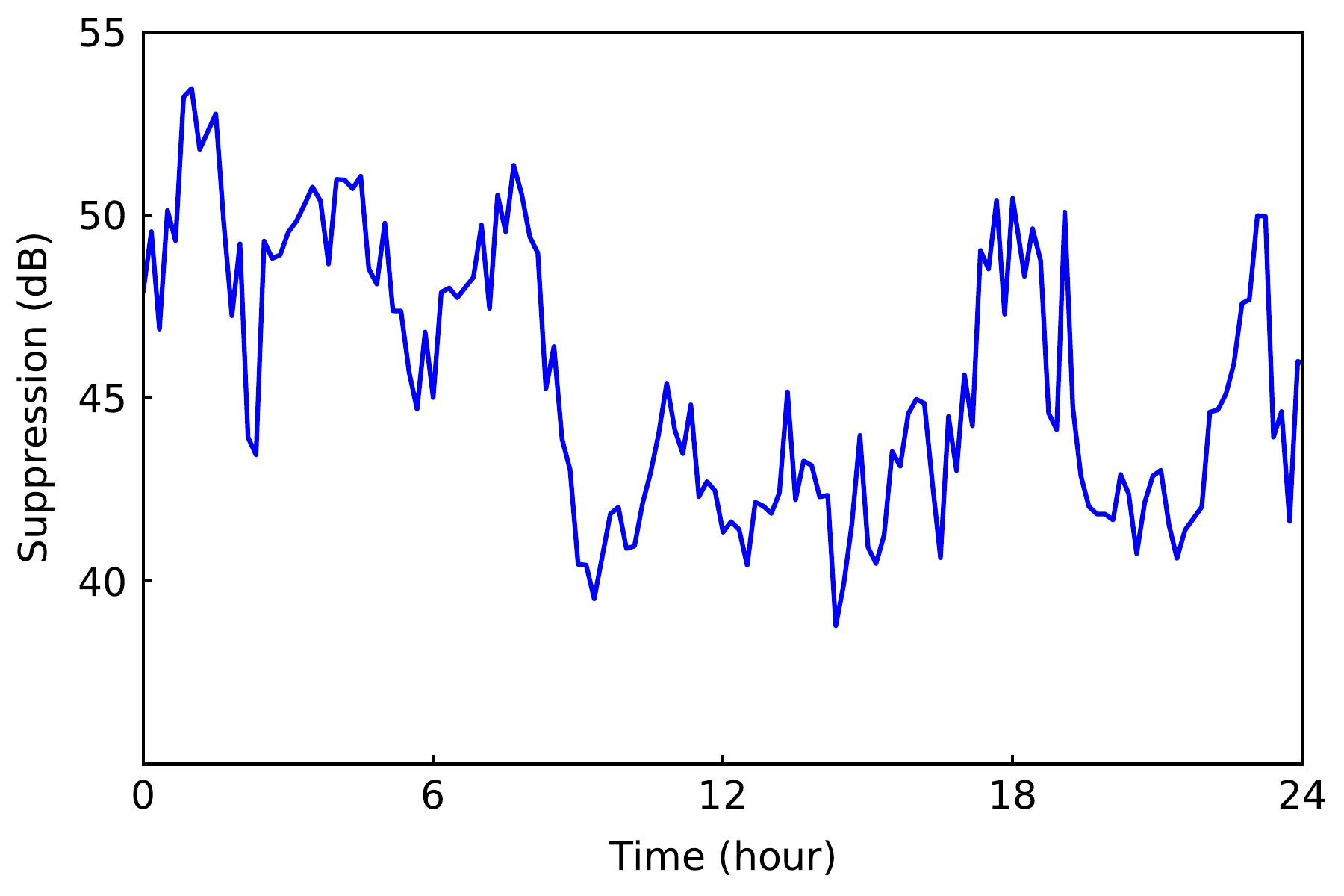}
	\caption{Feedforward suppression performance for injected phase modulation at 1\,MHz, recorded over a period of 24 hours.}
		\label{time_robust}%文中引用该图片代号
	\end{minipage}
	%\qquad
	%让图片换行
\end{figure}

While these results define the system's peak suppression capabilities, the open-loop nature of the feedforward design makes it inherently sensitive to environmental disturbances. To ensure long-term robustness, we stabilize both the beat-signal amplitude and the LO2 phase. Figure~\ref{time_robust} illustrates the stability of this approach, with suppression of the injected 1-MHz noise remaining consistently above 39\,dB over a 24-hour period. Residual fluctuations are likely caused by temperature variations and near-mode-hopping in the slave laser.

\begin{figure}[htbp]
\centering
\includegraphics[width=1\linewidth]{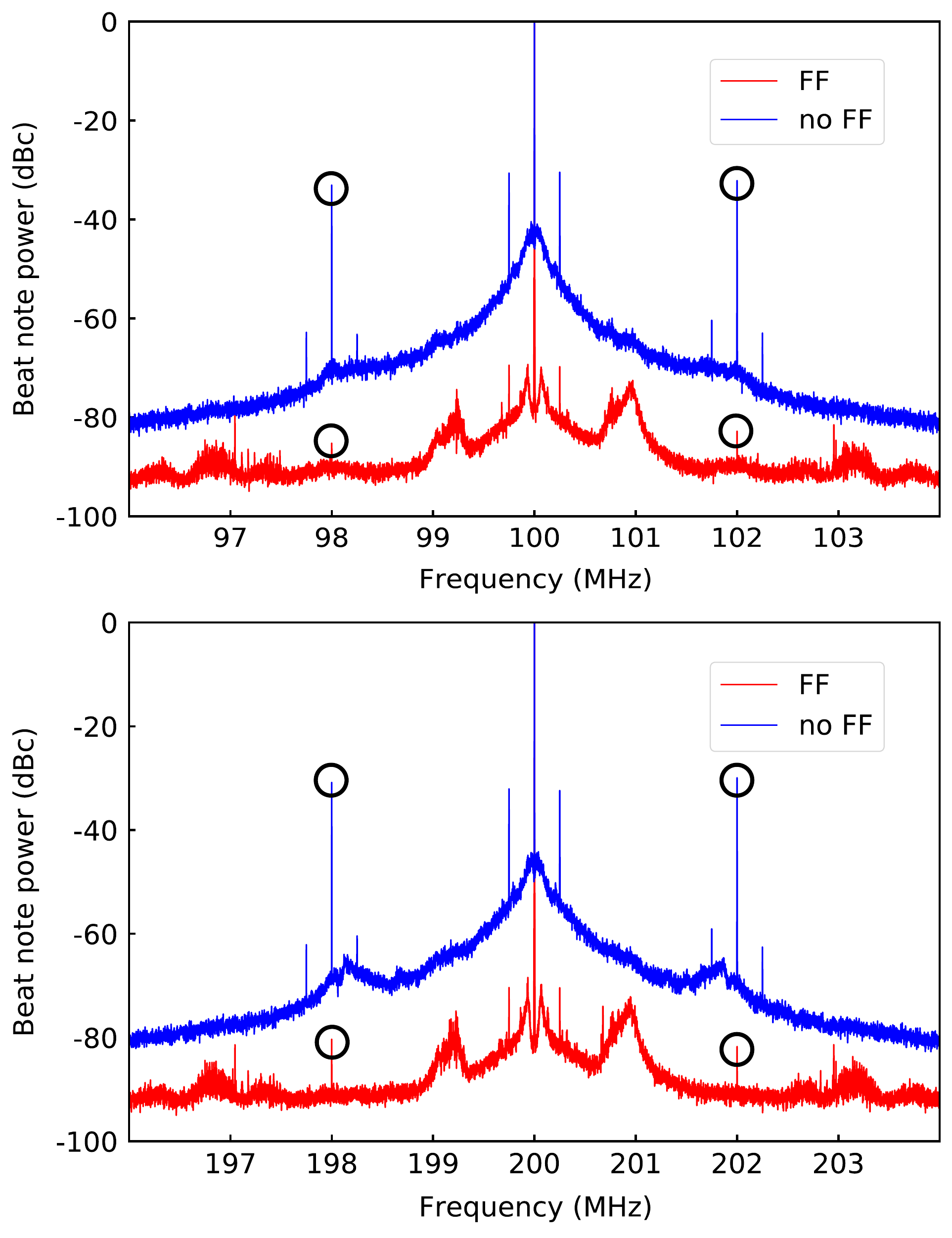} 
\caption{Performance of OPL feedforward at (a) $\Delta \omega=100$\,MHz, (b) $\Delta \omega=200$\,MHz, obtained with 500\,Hz resolution. 
}
\label{different_beat}
\end{figure}

Beyond temporal stability, a practical implementation very often requires the ability to tune the laser frequency without degrading performance. Our configuration facilitates easy adjustment of the master-slave offset, $\Delta\omega$, while maintaining high noise suppression. Changing $\Delta\omega$ simply requires proportional scaling of the LO1 and LO2 frequencies, which are linked by the PLL chip ratio. Figure~\ref{different_beat} compares the beat spectra with and without feedforward at (a) 100-MHz and (b) 200-MHz offsets; the resulting suppression is nearly identical to that observed at 240\,MHz [Fig.~\ref{suppression}(a)]. While we fine-tuned the P amplifier's gain by $\sim$1\% to keep the 2-MHz suppression above 50\,dB in these specific cases, the suppression remains above 37\,dB across the 100--200\,MHz range even without adjustment. This consistency is currently limited by the gain flatness of the electrical components. Nevertheless, the performance without recalibration suggests that, with automatic LO2 phase correction, the system could maintain high performance even during continuous ramping of $\Delta\omega$.

\section{Comparison of different feedforward methods}\label{sec:comparisons}
To date, feedforward phase-noise suppression techniques can be categorized by their phase-error detection mechanisms into three primary groups: delayed self-homodyne interferometry~\cite{2009_Bagheri_Semiconductor_laser_phase_noise_cancellation, 2012_Aflatouni_Wideband_tunable_laser_phase_noise_reduction, 2016_Okamoto_Toge_Manabe_Robust_Laser_Phase_Noise_Reduction,2019_HeZuyuan_AOM, 2017_Lintz_Note_MZI_FFFB, 2025_Leseleuc_FF}, Pound-Drever-Hall (PDH) detection using optical cavities~\cite{2024_Chao_PDHFeedForward,2025_Chao_Robust,2025_Hua_Feedforward_FDL}, and heterodyne beating with a reference laser~\cite{2010_Koke_NatPhys_AOM_FF_Comb,2012_Sala_Wide-bandwidth_phase_lock, 2015_Burkart_Optical_phase_cloning,2016_Watts_Phase_Noise_Reduction_of_an_Optical_Frequency_Comb_beat_MZM,2017_Besnard_beat_MZM,2020-Tan-Sensor-beat-FFFB-AOM, 2022_Li_Active_Cancellation_of_Servo-Induced_Noise,2024_Beat_IQmodulator}. Each approach entails specific performance trade-offs. Interferometric methods detect phase variations over a fixed time delay $\Delta T$ set by the arm-length imbalance. While the resulting signals are proportional to frequency noise at low Fourier frequencies ($f_n$), they undergo significant distortion when $f_n \ge 0.05/\Delta T$. Consequently, these systems suffer from poor low-frequency sensitivity and an architectural inability to provide high noise suppression over wide bandwidths.

In comparison, the PDH technique generates an error signal directly proportional to the instantaneous residual phase jitter when the laser is locked to the cavity, enabling suppression exceeding 30 dB over broad bandwidths~\cite{2024_Chao_PDHFeedForward,2025_Chao_Robust}. However, its performance is constrained at low frequencies by the cavity linewidth and at high frequencies by the modulation frequency, necessitating high-finesse cavities and high-frequency modulation~\cite{2024_Chao_PDHFeedForward}. As demonstrated in this work, the heterodyne beat-based scheme provides superior low-frequency suppression compared to alternative techniques. This approach can operate effectively down to the kilohertz range, with its high-frequency performance limited only by electronic bandwidth or beat frequency offset $\Delta\omega$, which easily allows for operation up to several tens of MHz.

\section{Summary}\label{sec:summary}
In summary, we have proposed and demonstrated a straightforward method for suppressing relative phase fluctuations between a pair of optically phase-locked lasers. This feedforward noise-suppression method derives the correction signal by recycling and demodulating the beat note between the phase-locked lasers. Using this approach, we achieve phase-noise suppression exceeding 30\,dB from 10\,kHz to 10\,MHz. By stabilizing the beat-signal amplitude and the demodulation local-oscillator phase, we ensure reliable performance over extended periods and across varying master-slave frequency offsets. Utilizing high-frequency photodetectors, mixers, and PLL chips, this method can readily extend to offsets of tens of GHz and support continuous frequency-offset ramping. This technique is well-suited for applications requiring high-precision coherent quantum control of atoms and molecules.

\begin{acknowledgments}
This work is supported by the National Natural Science Foundation of China (NSFC)(Grants 12234012.and No.W2431002),and the Quantum Science and Technology-National Science and Technology Major Project (2021ZD0302104).
\end{acknowledgments}

\newpage
\appendix
\section{Detailed setup}\label{section:detail set up}

\begin{figure}[htbp]
	\centering
	\begin{minipage}{1\linewidth}
		\centering
		\includegraphics[width=0.99\linewidth]{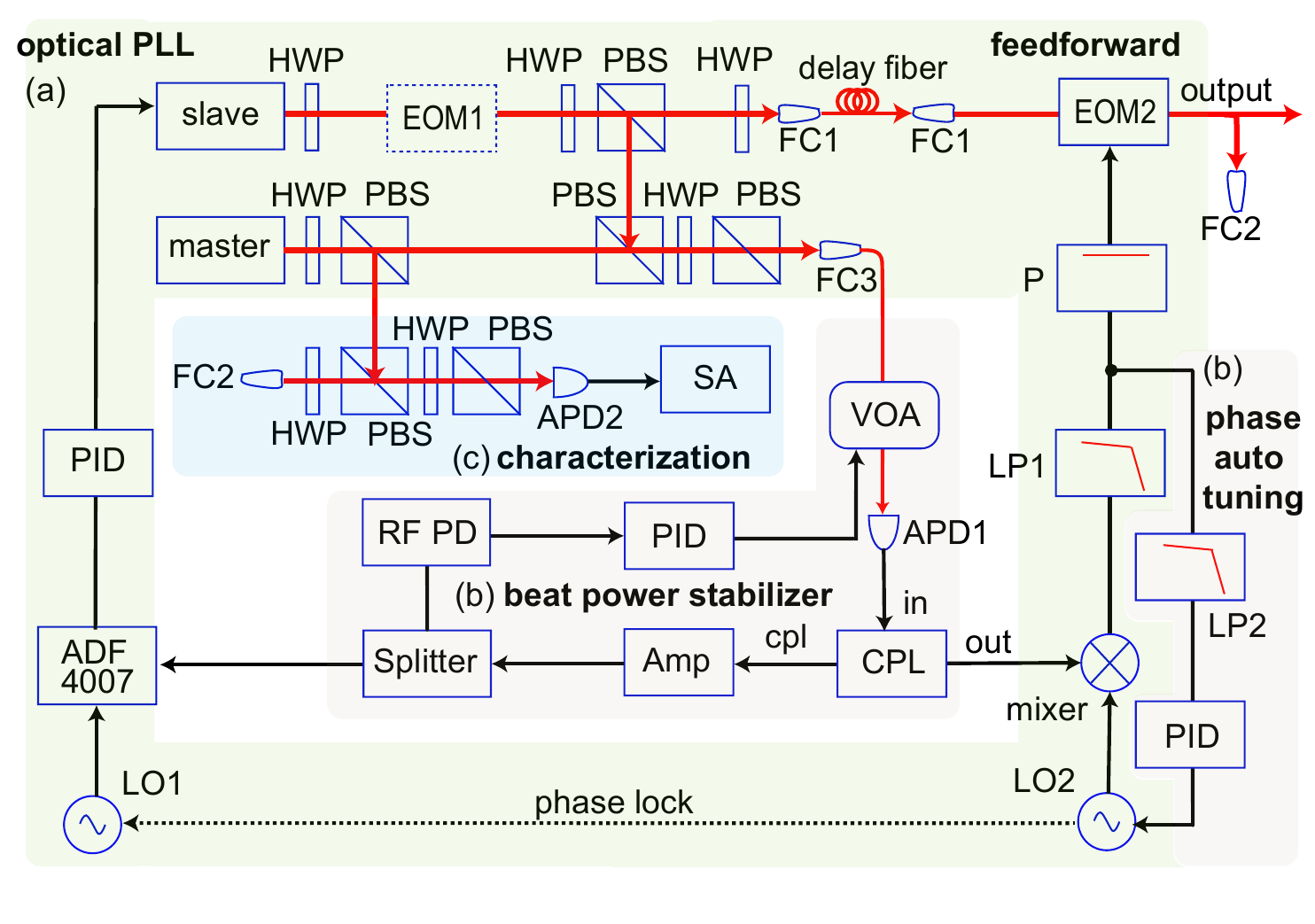}
	\caption{Detailed experimental setup. The green shaded area (a) encompasses the optical phase-locked loop and primary feedforward components. The beige shaded areas (b) include two feedback loops that enhance the robustness of the feedforward scheme. The blue shaded area (c) is used to characterize the feedforward effects. HWP, half-wave plate; PBS, polarizing beam splitter; FC, fiber coupler; VOA, variable optical attenuator; APD, avalanche photodiode; PID, proportional-integral-derivative controller; Splitter (Mini-Circuits ZFRSC-42-S+); RF PD, radio-frequency power detector (Mini-Circuits ZX47-40-S+); CPL, directional coupler (Mini-Circuits ZFDC-10-1+); Amp, RF amplifier (Mini-Circuits ZHL-72A+); SA, spectrum analyzer; LP, low-pass filter; P, home-built proportional amplifier.}
\label{fig:detailed_setup}
\end{minipage}
\end{figure}

Figure~\ref{fig:detailed_setup} provides a detailed view of our setup. The slave laser is a 1013-nm cat-eye semiconductor laser (MOGLabs). The master is a 1013-nm fiber laser (Preci-Laser FL-SF-1012-S). Their beat signal is detected by an avalanche photodiode (APD1, Thorlabs APD430C). Beat-signal amplitude stabilization is accomplished with a variable optical attenuator (Thorlabs V800PA). Optical phase locking is achieved using a phase discriminator (Analog Devices ADF4007). Residual phase fluctuations are extracted by demodulating the beat signal with a mixer (Mini-Circuits ZX05-1MHW-S+), followed by low-pass filtering (LP1, Mini-Circuits SLP-70+). The mixer's DC offset is monitored via a low-cutoff low-pass filter (LP2, LPF-10KK) and nullified by feedback to the LO2 output phase. Phase modulation is injected using a fiber EOM (EOM1, KEYANG PHOTONICS PM-10-10G). The feedforward correction is applied via another fiber EOM (EOM2, JENOPTIK PM1064). Feedforward performance is evaluated by detecting the post-EOM2 master-slave beat with APD2 and a spectrum analyzer (SA, Rohde \& Schwarz FSV3000).

\bibliographystyle{apsrev4-2}
% \bibliography{citation}
%\bibliography{citation}

\begin{thebibliography}{39}%
\makeatletter
\providecommand \@ifxundefined [1]{%
 \@ifx{#1\undefined}
}%
\providecommand \@ifnum [1]{%
 \ifnum #1\expandafter \@firstoftwo
 \else \expandafter \@secondoftwo
 \fi
}%
\providecommand \@ifx [1]{%
 \ifx #1\expandafter \@firstoftwo
 \else \expandafter \@secondoftwo
 \fi
}%
\providecommand \natexlab [1]{#1}%
\providecommand \enquote  [1]{``#1''}%
\providecommand \bibnamefont  [1]{#1}%
\providecommand \bibfnamefont [1]{#1}%
\providecommand \citenamefont [1]{#1}%
\providecommand \href@noop [0]{\@secondoftwo}%
\providecommand \href [0]{\begingroup \@sanitize@url \@href}%
\providecommand \@href[1]{\@@startlink{#1}\@@href}%
\providecommand \@@href[1]{\endgroup#1\@@endlink}%
\providecommand \@sanitize@url [0]{\catcode `\\12\catcode `\$12\catcode `\&12\catcode `\#12\catcode `\^12\catcode `\_12\catcode `\%12\relax}%
\providecommand \@@startlink[1]{}%
\providecommand \@@endlink[0]{}%
\providecommand \url  [0]{\begingroup\@sanitize@url \@url }%
\providecommand \@url [1]{\endgroup\@href {#1}{\urlprefix }}%
\providecommand \urlprefix  [0]{URL }%
\providecommand \Eprint [0]{\href }%
\providecommand \doibase [0]{https://doi.org/}%
\providecommand \selectlanguage [0]{\@gobble}%
\providecommand \bibinfo  [0]{\@secondoftwo}%
\providecommand \bibfield  [0]{\@secondoftwo}%
\providecommand \translation [1]{[#1]}%
\providecommand \BibitemOpen [0]{}%
\providecommand \bibitemStop [0]{}%
\providecommand \bibitemNoStop [0]{.\EOS\space}%
\providecommand \EOS [0]{\spacefactor3000\relax}%
\providecommand \BibitemShut  [1]{\csname bibitem#1\endcsname}%
\let\auto@bib@innerbib\@empty
%</preamble>
\bibitem [{\citenamefont {Hassan}\ \emph {et~al.}(2025)\citenamefont {Hassan}, \citenamefont {Beloy}, \citenamefont {Siegel}, \citenamefont {Kobayashi}, \citenamefont {Swiler}, \citenamefont {Grogan}, \citenamefont {Brown}, \citenamefont {Rojo}, \citenamefont {Bothwell}, \citenamefont {Hunt}, \citenamefont {Halaoui},\ and\ \citenamefont {Ludlow}}]{2025_Hassan_Beloy_Siegel_Kobayashi_Swiler_Grogan_Brown_Rojo_Bothwell_Hunt_Cryogenic_Optical_Lattice_Clock}%
  \BibitemOpen
  \bibfield  {author} {\bibinfo {author} {\bibfnamefont {Y.~S.}\ \bibnamefont {Hassan}}, \bibinfo {author} {\bibfnamefont {K.}~\bibnamefont {Beloy}}, \bibinfo {author} {\bibfnamefont {J.~L.}\ \bibnamefont {Siegel}}, \bibinfo {author} {\bibfnamefont {T.}~\bibnamefont {Kobayashi}}, \bibinfo {author} {\bibfnamefont {E.}~\bibnamefont {Swiler}}, \bibinfo {author} {\bibfnamefont {T.}~\bibnamefont {Grogan}}, \bibinfo {author} {\bibfnamefont {R.~C.}\ \bibnamefont {Brown}}, \bibinfo {author} {\bibfnamefont {T.}~\bibnamefont {Rojo}}, \bibinfo {author} {\bibfnamefont {T.}~\bibnamefont {Bothwell}}, \bibinfo {author} {\bibfnamefont {B.~D.}\ \bibnamefont {Hunt}}, \bibinfo {author} {\bibfnamefont {A.}~\bibnamefont {Halaoui}},\ and\ \bibinfo {author} {\bibfnamefont {A.~D.}\ \bibnamefont {Ludlow}},\ }\href {https://doi.org/10.1103/4tky-jmsm} {\bibfield  {journal} {\bibinfo  {journal} {Phys. Rev. Lett.}\ }\textbf {\bibinfo {volume} {135}},\ \bibinfo {pages} {063402} (\bibinfo {year} {2025})}\BibitemShut {NoStop}%
\bibitem [{\citenamefont {{The LIGO Scientific Collaboration et al}}(2015)}]{2015_LIGO_AdvancedLIGO}%
  \BibitemOpen
  \bibfield  {author} {\bibinfo {author} {\bibnamefont {{The LIGO Scientific Collaboration et al}}},\ }\href {https://doi.org/10.1088/0264-9381/32/7/074001} {\bibfield  {journal} {\bibinfo  {journal} {Classical and Quantum Gravity}\ }\textbf {\bibinfo {volume} {32}},\ \bibinfo {pages} {074001} (\bibinfo {year} {2015})}\BibitemShut {NoStop}%
\bibitem [{\citenamefont {Kolkowitz}\ \emph {et~al.}(2016)\citenamefont {Kolkowitz}, \citenamefont {Pikovski}, \citenamefont {Langellier}, \citenamefont {Lukin}, \citenamefont {Walsworth},\ and\ \citenamefont {Ye}}]{2016_Ye_GWD_with_atomic_clocks}%
  \BibitemOpen
  \bibfield  {author} {\bibinfo {author} {\bibfnamefont {S.}~\bibnamefont {Kolkowitz}}, \bibinfo {author} {\bibfnamefont {I.}~\bibnamefont {Pikovski}}, \bibinfo {author} {\bibfnamefont {N.}~\bibnamefont {Langellier}}, \bibinfo {author} {\bibfnamefont {M.~D.}\ \bibnamefont {Lukin}}, \bibinfo {author} {\bibfnamefont {R.~L.}\ \bibnamefont {Walsworth}},\ and\ \bibinfo {author} {\bibfnamefont {J.}~\bibnamefont {Ye}},\ }\href {https://doi.org/10.1103/PhysRevD.94.124043} {\bibfield  {journal} {\bibinfo  {journal} {Phys. Rev. D}\ }\textbf {\bibinfo {volume} {94}},\ \bibinfo {pages} {124043} (\bibinfo {year} {2016})}\BibitemShut {NoStop}%
\bibitem [{\citenamefont {Djevahirdjian}\ \emph {et~al.}(2023)\citenamefont {Djevahirdjian}, \citenamefont {Lechevallier}, \citenamefont {Martin-Drumel}, \citenamefont {Pirali}, \citenamefont {Ducournau}, \citenamefont {Kassi},\ and\ \citenamefont {Kassi}}]{2023_Djevahirdjian_Lechevallier_Martin-Drumel_Pirali_Ducournau_Kassi_Frequency_stable_and_low_phase_noise_THz}%
  \BibitemOpen
  \bibfield  {author} {\bibinfo {author} {\bibfnamefont {L.}~\bibnamefont {Djevahirdjian}}, \bibinfo {author} {\bibfnamefont {L.}~\bibnamefont {Lechevallier}}, \bibinfo {author} {\bibfnamefont {M.-A.}\ \bibnamefont {Martin-Drumel}}, \bibinfo {author} {\bibfnamefont {O.}~\bibnamefont {Pirali}}, \bibinfo {author} {\bibfnamefont {G.}~\bibnamefont {Ducournau}}, \bibinfo {author} {\bibfnamefont {R.}~\bibnamefont {Kassi}},\ and\ \bibinfo {author} {\bibfnamefont {S.}~\bibnamefont {Kassi}},\ }\href@noop {} {\bibfield  {journal} {\bibinfo  {journal} {Nature Communications}\ }\textbf {\bibinfo {volume} {14}},\ \bibinfo {pages} {7162} (\bibinfo {year} {2023})}\BibitemShut {NoStop}%
\bibitem [{\citenamefont {He}\ \emph {et~al.}(2025)\citenamefont {He}, \citenamefont {Yang}, \citenamefont {Meng}, \citenamefont {Yu}, \citenamefont {Zhang}, \citenamefont {Yang}, \citenamefont {Zuo}, \citenamefont {Lin}, \citenamefont {Chen}, \citenamefont {Fang},\ and\ \citenamefont {Xie}}]{2025_He_NatComm_microwave_generation}%
  \BibitemOpen
  \bibfield  {author} {\bibinfo {author} {\bibfnamefont {B.}~\bibnamefont {He}}, \bibinfo {author} {\bibfnamefont {J.}~\bibnamefont {Yang}}, \bibinfo {author} {\bibfnamefont {F.}~\bibnamefont {Meng}}, \bibinfo {author} {\bibfnamefont {J.}~\bibnamefont {Yu}}, \bibinfo {author} {\bibfnamefont {C.}~\bibnamefont {Zhang}}, \bibinfo {author} {\bibfnamefont {Q.-F.}\ \bibnamefont {Yang}}, \bibinfo {author} {\bibfnamefont {Y.}~\bibnamefont {Zuo}}, \bibinfo {author} {\bibfnamefont {Y.}~\bibnamefont {Lin}}, \bibinfo {author} {\bibfnamefont {Z.}~\bibnamefont {Chen}}, \bibinfo {author} {\bibfnamefont {Z.}~\bibnamefont {Fang}},\ and\ \bibinfo {author} {\bibfnamefont {X.}~\bibnamefont {Xie}},\ }\href {https://doi.org/10.1038/s41467-025-59401-1} {\bibfield  {journal} {\bibinfo  {journal} {Nature Communications}\ }\textbf {\bibinfo {volume} {16}},\ \bibinfo {pages} {4034} (\bibinfo {year} {2025})}\BibitemShut {NoStop}%
\bibitem [{\citenamefont {Egan}\ \emph {et~al.}(2021)\citenamefont {Egan}, \citenamefont {Debroy}, \citenamefont {Noel}, \citenamefont {Risinger}, \citenamefont {Zhu}, \citenamefont {Biswas}, \citenamefont {Newman}, \citenamefont {Li}, \citenamefont {Brown}, \citenamefont {Cetina},\ and\ \citenamefont {Monroe}}]{2021_Egan_Debroy_Noel_Risinger_Zhu_Biswas_Newman_Li_Brown_Cetina_Fault-tolerant_control}%
  \BibitemOpen
  \bibfield  {author} {\bibinfo {author} {\bibfnamefont {L.}~\bibnamefont {Egan}}, \bibinfo {author} {\bibfnamefont {D.~M.}\ \bibnamefont {Debroy}}, \bibinfo {author} {\bibfnamefont {C.}~\bibnamefont {Noel}}, \bibinfo {author} {\bibfnamefont {A.}~\bibnamefont {Risinger}}, \bibinfo {author} {\bibfnamefont {D.}~\bibnamefont {Zhu}}, \bibinfo {author} {\bibfnamefont {D.}~\bibnamefont {Biswas}}, \bibinfo {author} {\bibfnamefont {M.}~\bibnamefont {Newman}}, \bibinfo {author} {\bibfnamefont {M.}~\bibnamefont {Li}}, \bibinfo {author} {\bibfnamefont {K.~R.}\ \bibnamefont {Brown}}, \bibinfo {author} {\bibfnamefont {M.}~\bibnamefont {Cetina}},\ and\ \bibinfo {author} {\bibfnamefont {C.}~\bibnamefont {Monroe}},\ }\href {https://doi.org/10.1038/s41586-021-03928-y} {\bibfield  {journal} {\bibinfo  {journal} {Nature}\ }\textbf {\bibinfo {volume} {598}},\ \bibinfo {pages} {281–286} (\bibinfo {year} {2021})}\BibitemShut {NoStop}%
\bibitem [{\citenamefont {Smith}\ \emph {et~al.}(2025)\citenamefont {Smith}, \citenamefont {Leu}, \citenamefont {Miyanishi}, \citenamefont {Gely},\ and\ \citenamefont {Lucas}}]{2025_Smith_Single-Qubit_Gates_with_Errors_at_10-7}%
  \BibitemOpen
  \bibfield  {author} {\bibinfo {author} {\bibfnamefont {M.~C.}\ \bibnamefont {Smith}}, \bibinfo {author} {\bibfnamefont {A.~D.}\ \bibnamefont {Leu}}, \bibinfo {author} {\bibfnamefont {K.}~\bibnamefont {Miyanishi}}, \bibinfo {author} {\bibfnamefont {M.~F.}\ \bibnamefont {Gely}},\ and\ \bibinfo {author} {\bibfnamefont {D.~M.}\ \bibnamefont {Lucas}},\ }\href {https://doi.org/10.1103/42w2-6ccy} {\bibfield  {journal} {\bibinfo  {journal} {Phys. Rev. Lett.}\ }\textbf {\bibinfo {volume} {134}},\ \bibinfo {pages} {230601} (\bibinfo {year} {2025})}\BibitemShut {NoStop}%
\bibitem [{\citenamefont {Yan}\ \emph {et~al.}(2025)\citenamefont {Yan}, \citenamefont {Lannig}, \citenamefont {Milner}, \citenamefont {Frankel}, \citenamefont {Lewis}, \citenamefont {Lee}, \citenamefont {Kim},\ and\ \citenamefont {Ye}}]{2025_Ye_PRX_High_Power_Clock_Laser}%
  \BibitemOpen
  \bibfield  {author} {\bibinfo {author} {\bibfnamefont {L.}~\bibnamefont {Yan}}, \bibinfo {author} {\bibfnamefont {S.}~\bibnamefont {Lannig}}, \bibinfo {author} {\bibfnamefont {W.~R.}\ \bibnamefont {Milner}}, \bibinfo {author} {\bibfnamefont {M.~N.}\ \bibnamefont {Frankel}}, \bibinfo {author} {\bibfnamefont {B.}~\bibnamefont {Lewis}}, \bibinfo {author} {\bibfnamefont {D.}~\bibnamefont {Lee}}, \bibinfo {author} {\bibfnamefont {K.}~\bibnamefont {Kim}},\ and\ \bibinfo {author} {\bibfnamefont {J.}~\bibnamefont {Ye}},\ }\href {https://doi.org/10.1103/qw53-8b8r} {\bibfield  {journal} {\bibinfo  {journal} {Phys. Rev. X}\ }\textbf {\bibinfo {volume} {15}},\ \bibinfo {pages} {031055} (\bibinfo {year} {2025})}\BibitemShut {NoStop}%
\bibitem [{\citenamefont {Evered}\ \emph {et~al.}(2023)\citenamefont {Evered}, \citenamefont {Bluvstein}, \citenamefont {Kalinowski}, \citenamefont {Ebadi}, \citenamefont {Manovitz}, \citenamefont {Zhou}, \citenamefont {Li}, \citenamefont {Geim}, \citenamefont {Wang}, \citenamefont {Maskara}, \citenamefont {Levine}, \citenamefont {Semeghini}, \citenamefont {Greiner}, \citenamefont {Vuleti{\'c}},\ and\ \citenamefont {Lukin}}]{2023_Evered_High-fidelity_parallel_entangling_gates}%
  \BibitemOpen
  \bibfield  {author} {\bibinfo {author} {\bibfnamefont {S.~J.}\ \bibnamefont {Evered}}, \bibinfo {author} {\bibfnamefont {D.}~\bibnamefont {Bluvstein}}, \bibinfo {author} {\bibfnamefont {M.}~\bibnamefont {Kalinowski}}, \bibinfo {author} {\bibfnamefont {S.}~\bibnamefont {Ebadi}}, \bibinfo {author} {\bibfnamefont {T.}~\bibnamefont {Manovitz}}, \bibinfo {author} {\bibfnamefont {H.}~\bibnamefont {Zhou}}, \bibinfo {author} {\bibfnamefont {S.~H.}\ \bibnamefont {Li}}, \bibinfo {author} {\bibfnamefont {A.~A.}\ \bibnamefont {Geim}}, \bibinfo {author} {\bibfnamefont {T.~T.}\ \bibnamefont {Wang}}, \bibinfo {author} {\bibfnamefont {N.}~\bibnamefont {Maskara}}, \bibinfo {author} {\bibfnamefont {H.}~\bibnamefont {Levine}}, \bibinfo {author} {\bibfnamefont {G.}~\bibnamefont {Semeghini}}, \bibinfo {author} {\bibfnamefont {M.}~\bibnamefont {Greiner}}, \bibinfo {author} {\bibfnamefont {V.}~\bibnamefont {Vuleti{\'c}}},\ and\ \bibinfo {author} {\bibfnamefont {M.~D.}\ \bibnamefont {Lukin}},\ }\href@noop {} {\bibfield  {journal}
  {\bibinfo  {journal} {Nature}\ }\textbf {\bibinfo {volume} {622}},\ \bibinfo {pages} {268} (\bibinfo {year} {2023})}\BibitemShut {NoStop}%
\bibitem [{\citenamefont {Bluvstein}\ \emph {et~al.}(2024)\citenamefont {Bluvstein}, \citenamefont {Evered}, \citenamefont {Geim}, \citenamefont {Li}, \citenamefont {Zhou}, \citenamefont {Manovitz}, \citenamefont {Ebadi}, \citenamefont {Cain}, \citenamefont {Kalinowski}, \citenamefont {Hangleiter}, \citenamefont {Bonilla~Ataides}, \citenamefont {Maskara}, \citenamefont {Cong}, \citenamefont {Gao}, \citenamefont {Sales~Rodriguez}, \citenamefont {Karolyshyn}, \citenamefont {Semeghini}, \citenamefont {Gullans}, \citenamefont {Greiner}, \citenamefont {Vuleti{\'c}},\ and\ \citenamefont {Lukin}}]{2024_Bluvstein_Logical_quantum_processor_based}%
  \BibitemOpen
  \bibfield  {author} {\bibinfo {author} {\bibfnamefont {D.}~\bibnamefont {Bluvstein}}, \bibinfo {author} {\bibfnamefont {S.~J.}\ \bibnamefont {Evered}}, \bibinfo {author} {\bibfnamefont {A.~A.}\ \bibnamefont {Geim}}, \bibinfo {author} {\bibfnamefont {S.~H.}\ \bibnamefont {Li}}, \bibinfo {author} {\bibfnamefont {H.}~\bibnamefont {Zhou}}, \bibinfo {author} {\bibfnamefont {T.}~\bibnamefont {Manovitz}}, \bibinfo {author} {\bibfnamefont {S.}~\bibnamefont {Ebadi}}, \bibinfo {author} {\bibfnamefont {M.}~\bibnamefont {Cain}}, \bibinfo {author} {\bibfnamefont {M.}~\bibnamefont {Kalinowski}}, \bibinfo {author} {\bibfnamefont {D.}~\bibnamefont {Hangleiter}}, \bibinfo {author} {\bibfnamefont {J.~P.}\ \bibnamefont {Bonilla~Ataides}}, \bibinfo {author} {\bibfnamefont {N.}~\bibnamefont {Maskara}}, \bibinfo {author} {\bibfnamefont {I.}~\bibnamefont {Cong}}, \bibinfo {author} {\bibfnamefont {X.}~\bibnamefont {Gao}}, \bibinfo {author} {\bibfnamefont {P.}~\bibnamefont {Sales~Rodriguez}}, \bibinfo {author} {\bibfnamefont
  {T.}~\bibnamefont {Karolyshyn}}, \bibinfo {author} {\bibfnamefont {G.}~\bibnamefont {Semeghini}}, \bibinfo {author} {\bibfnamefont {M.~J.}\ \bibnamefont {Gullans}}, \bibinfo {author} {\bibfnamefont {M.}~\bibnamefont {Greiner}}, \bibinfo {author} {\bibfnamefont {V.}~\bibnamefont {Vuleti{\'c}}},\ and\ \bibinfo {author} {\bibfnamefont {M.~D.}\ \bibnamefont {Lukin}},\ }\href@noop {} {\bibfield  {journal} {\bibinfo  {journal} {Nature}\ }\textbf {\bibinfo {volume} {626}},\ \bibinfo {pages} {58} (\bibinfo {year} {2024})}\BibitemShut {NoStop}%
\bibitem [{\citenamefont {Fu}\ \emph {et~al.}(2022)\citenamefont {Fu}, \citenamefont {Xu}, \citenamefont {Sun}, \citenamefont {Liu}, \citenamefont {He}, \citenamefont {Li}, \citenamefont {Liu}, \citenamefont {Li}, \citenamefont {Wang}, \citenamefont {Liu},\ and\ \citenamefont {Zhan}}]{2022_Zhan_Rydberg_gate}%
  \BibitemOpen
  \bibfield  {author} {\bibinfo {author} {\bibfnamefont {Z.}~\bibnamefont {Fu}}, \bibinfo {author} {\bibfnamefont {P.}~\bibnamefont {Xu}}, \bibinfo {author} {\bibfnamefont {Y.}~\bibnamefont {Sun}}, \bibinfo {author} {\bibfnamefont {Y.-Y.}\ \bibnamefont {Liu}}, \bibinfo {author} {\bibfnamefont {X.-D.}\ \bibnamefont {He}}, \bibinfo {author} {\bibfnamefont {X.}~\bibnamefont {Li}}, \bibinfo {author} {\bibfnamefont {M.}~\bibnamefont {Liu}}, \bibinfo {author} {\bibfnamefont {R.-B.}\ \bibnamefont {Li}}, \bibinfo {author} {\bibfnamefont {J.}~\bibnamefont {Wang}}, \bibinfo {author} {\bibfnamefont {L.}~\bibnamefont {Liu}},\ and\ \bibinfo {author} {\bibfnamefont {M.-S.}\ \bibnamefont {Zhan}},\ }\href {https://doi.org/10.1103/PhysRevA.105.042430} {\bibfield  {journal} {\bibinfo  {journal} {Phys. Rev. A}\ }\textbf {\bibinfo {volume} {105}},\ \bibinfo {pages} {042430} (\bibinfo {year} {2022})}\BibitemShut {NoStop}%
\bibitem [{\citenamefont {Molony}\ \emph {et~al.}(2014)\citenamefont {Molony}, \citenamefont {Gregory}, \citenamefont {Ji}, \citenamefont {Lu}, \citenamefont {K\"oppinger}, \citenamefont {Le~Sueur}, \citenamefont {Blackley}, \citenamefont {Hutson},\ and\ \citenamefont {Cornish}}]{2014_Molony_Creation_of_Ultracold}%
  \BibitemOpen
  \bibfield  {author} {\bibinfo {author} {\bibfnamefont {P.~K.}\ \bibnamefont {Molony}}, \bibinfo {author} {\bibfnamefont {P.~D.}\ \bibnamefont {Gregory}}, \bibinfo {author} {\bibfnamefont {Z.}~\bibnamefont {Ji}}, \bibinfo {author} {\bibfnamefont {B.}~\bibnamefont {Lu}}, \bibinfo {author} {\bibfnamefont {M.~P.}\ \bibnamefont {K\"oppinger}}, \bibinfo {author} {\bibfnamefont {C.~R.}\ \bibnamefont {Le~Sueur}}, \bibinfo {author} {\bibfnamefont {C.~L.}\ \bibnamefont {Blackley}}, \bibinfo {author} {\bibfnamefont {J.~M.}\ \bibnamefont {Hutson}},\ and\ \bibinfo {author} {\bibfnamefont {S.~L.}\ \bibnamefont {Cornish}},\ }\href {https://doi.org/10.1103/PhysRevLett.113.255301} {\bibfield  {journal} {\bibinfo  {journal} {Phys. Rev. Lett.}\ }\textbf {\bibinfo {volume} {113}},\ \bibinfo {pages} {255301} (\bibinfo {year} {2014})}\BibitemShut {NoStop}%
\bibitem [{\citenamefont {Matei}\ \emph {et~al.}(2017)\citenamefont {Matei}, \citenamefont {Legero}, \citenamefont {H\"afner}, \citenamefont {Grebing}, \citenamefont {Weyrich}, \citenamefont {Zhang}, \citenamefont {Sonderhouse}, \citenamefont {Robinson}, \citenamefont {Ye}, \citenamefont {Riehle},\ and\ \citenamefont {Sterr}}]{2017_YeJun_mHzlaser}%
  \BibitemOpen
  \bibfield  {author} {\bibinfo {author} {\bibfnamefont {D.~G.}\ \bibnamefont {Matei}}, \bibinfo {author} {\bibfnamefont {T.}~\bibnamefont {Legero}}, \bibinfo {author} {\bibfnamefont {S.}~\bibnamefont {H\"afner}}, \bibinfo {author} {\bibfnamefont {C.}~\bibnamefont {Grebing}}, \bibinfo {author} {\bibfnamefont {R.}~\bibnamefont {Weyrich}}, \bibinfo {author} {\bibfnamefont {W.}~\bibnamefont {Zhang}}, \bibinfo {author} {\bibfnamefont {L.}~\bibnamefont {Sonderhouse}}, \bibinfo {author} {\bibfnamefont {J.~M.}\ \bibnamefont {Robinson}}, \bibinfo {author} {\bibfnamefont {J.}~\bibnamefont {Ye}}, \bibinfo {author} {\bibfnamefont {F.}~\bibnamefont {Riehle}},\ and\ \bibinfo {author} {\bibfnamefont {U.}~\bibnamefont {Sterr}},\ }\href {https://doi.org/10.1103/PhysRevLett.118.263202} {\bibfield  {journal} {\bibinfo  {journal} {Phys. Rev. Lett.}\ }\textbf {\bibinfo {volume} {118}},\ \bibinfo {pages} {263202} (\bibinfo {year} {2017})}\BibitemShut {NoStop}%
\bibitem [{\citenamefont {Chen}\ \emph {et~al.}(2025)\citenamefont {Chen}, \citenamefont {Zeng}, \citenamefont {Wang}, \citenamefont {Zhang}, \citenamefont {Lei}, \citenamefont {Li}, \citenamefont {Pang}, \citenamefont {Huang},\ and\ \citenamefont {Zhang}}]{2025_Xibo_narrowlinewidthcavity}%
  \BibitemOpen
  \bibfield  {author} {\bibinfo {author} {\bibfnamefont {Z.-A.}\ \bibnamefont {Chen}}, \bibinfo {author} {\bibfnamefont {H.-R.}\ \bibnamefont {Zeng}}, \bibinfo {author} {\bibfnamefont {W.-W.}\ \bibnamefont {Wang}}, \bibinfo {author} {\bibfnamefont {H.}~\bibnamefont {Zhang}}, \bibinfo {author} {\bibfnamefont {R.-Q.}\ \bibnamefont {Lei}}, \bibinfo {author} {\bibfnamefont {J.-Z.}\ \bibnamefont {Li}}, \bibinfo {author} {\bibfnamefont {C.-Y.}\ \bibnamefont {Pang}}, \bibinfo {author} {\bibfnamefont {S.-S.}\ \bibnamefont {Huang}},\ and\ \bibinfo {author} {\bibfnamefont {X.}~\bibnamefont {Zhang}},\ }\href {https://doi.org/https://doi.org/10.1016/j.scib.2025.08.050} {\bibfield  {journal} {\bibinfo  {journal} {Science Bulletin}\ }\textbf {\bibinfo {volume} {70}},\ \bibinfo {pages} {3337} (\bibinfo {year} {2025})}\BibitemShut {NoStop}%
\bibitem [{\citenamefont {Rabinowitz}\ \emph {et~al.}(1963)\citenamefont {Rabinowitz}, \citenamefont {Latourrette},\ and\ \citenamefont {Gould}}]{1963_Rabinowitz_AFC_Optical_Heterodyne_Detector}%
  \BibitemOpen
  \bibfield  {author} {\bibinfo {author} {\bibfnamefont {P.}~\bibnamefont {Rabinowitz}}, \bibinfo {author} {\bibfnamefont {J.}~\bibnamefont {Latourrette}},\ and\ \bibinfo {author} {\bibfnamefont {G.}~\bibnamefont {Gould}},\ }\href {https://doi.org/10.1109/proc.1963.2292} {\bibfield  {journal} {\bibinfo  {journal} {Proceedings of the IEEE}\ }\textbf {\bibinfo {volume} {51}},\ \bibinfo {pages} {857–858} (\bibinfo {year} {1963})}\BibitemShut {NoStop}%
\bibitem [{\citenamefont {Enloe}\ and\ \citenamefont {Rodda}(1965)}]{1965_Enloe_Laser_phase-locked_loop}%
  \BibitemOpen
  \bibfield  {author} {\bibinfo {author} {\bibfnamefont {L.}~\bibnamefont {Enloe}}\ and\ \bibinfo {author} {\bibfnamefont {J.}~\bibnamefont {Rodda}},\ }\href {https://doi.org/10.1109/PROC.1965.3585} {\bibfield  {journal} {\bibinfo  {journal} {Proceedings of the IEEE}\ }\textbf {\bibinfo {volume} {53}},\ \bibinfo {pages} {165} (\bibinfo {year} {1965})}\BibitemShut {NoStop}%
\bibitem [{\citenamefont {Steele}(1983)}]{1983_Steele}%
  \BibitemOpen
  \bibfield  {author} {\bibinfo {author} {\bibfnamefont {R.}~\bibnamefont {Steele}},\ }\href {https://doi.org/10.1049/el:19830051} {\bibfield  {journal} {\bibinfo  {journal} {Electronics Letters}\ }\textbf {\bibinfo {volume} {19}},\ \bibinfo {pages} {69–71} (\bibinfo {year} {1983})}\BibitemShut {NoStop}%
\bibitem [{\citenamefont {Hall}\ \emph {et~al.}(1987)\citenamefont {Hall}, \citenamefont {Long-Sheng},\ and\ \citenamefont {Kramer}}]{1987_Hall_Principles_of_optical_phase-locking}%
  \BibitemOpen
  \bibfield  {author} {\bibinfo {author} {\bibfnamefont {J.}~\bibnamefont {Hall}}, \bibinfo {author} {\bibfnamefont {M.}~\bibnamefont {Long-Sheng}},\ and\ \bibinfo {author} {\bibfnamefont {G.}~\bibnamefont {Kramer}},\ }\href {https://doi.org/10.1109/jqe.1987.1073354} {\bibfield  {journal} {\bibinfo  {journal} {IEEE Journal of Quantum Electronics}\ }\textbf {\bibinfo {volume} {23}},\ \bibinfo {pages} {427–437} (\bibinfo {year} {1987})}\BibitemShut {NoStop}%
\bibitem [{\citenamefont {Ye}\ and\ \citenamefont {Hall}(1999)}]{1999_Ye_Hall_Optical_phase_locking_in_the_microradian_domain}%
  \BibitemOpen
  \bibfield  {author} {\bibinfo {author} {\bibfnamefont {J.}~\bibnamefont {Ye}}\ and\ \bibinfo {author} {\bibfnamefont {J.~L.}\ \bibnamefont {Hall}},\ }\href {https://doi.org/10.1364/ol.24.001838} {\bibfield  {journal} {\bibinfo  {journal} {Optics Letters}\ }\textbf {\bibinfo {volume} {24}},\ \bibinfo {pages} {1838} (\bibinfo {year} {1999})}\BibitemShut {NoStop}%
\bibitem [{\citenamefont {Guionie}\ \emph {et~al.}(2018)\citenamefont {Guionie}, \citenamefont {Frein}, \citenamefont {Carr\'{e}}, \citenamefont {Loas}, \citenamefont {Bondu}, \citenamefont {Pinsard}, \citenamefont {Lablonde}, \citenamefont {Cadier}, \citenamefont {Alouini}, \citenamefont {Romanelli}, \citenamefont {Vallet},\ and\ \citenamefont {Brunel}}]{2018_Brunel_subHz_PLL}%
  \BibitemOpen
  \bibfield  {author} {\bibinfo {author} {\bibfnamefont {M.}~\bibnamefont {Guionie}}, \bibinfo {author} {\bibfnamefont {L.}~\bibnamefont {Frein}}, \bibinfo {author} {\bibfnamefont {A.}~\bibnamefont {Carr\'{e}}}, \bibinfo {author} {\bibfnamefont {G.}~\bibnamefont {Loas}}, \bibinfo {author} {\bibfnamefont {F.}~\bibnamefont {Bondu}}, \bibinfo {author} {\bibfnamefont {E.}~\bibnamefont {Pinsard}}, \bibinfo {author} {\bibfnamefont {L.}~\bibnamefont {Lablonde}}, \bibinfo {author} {\bibfnamefont {B.}~\bibnamefont {Cadier}}, \bibinfo {author} {\bibfnamefont {M.}~\bibnamefont {Alouini}}, \bibinfo {author} {\bibfnamefont {M.}~\bibnamefont {Romanelli}}, \bibinfo {author} {\bibfnamefont {M.}~\bibnamefont {Vallet}},\ and\ \bibinfo {author} {\bibfnamefont {M.}~\bibnamefont {Brunel}},\ }\href {https://doi.org/10.1364/OE.26.003483} {\bibfield  {journal} {\bibinfo  {journal} {Opt. Express}\ }\textbf {\bibinfo {volume} {26}},\ \bibinfo {pages} {3483} (\bibinfo {year} {2018})}\BibitemShut {NoStop}%
\bibitem [{\citenamefont {Bagheri}\ \emph {et~al.}(2009)\citenamefont {Bagheri}, \citenamefont {Aflatouni}, \citenamefont {Imani}, \citenamefont {Goel},\ and\ \citenamefont {Hashemi}}]{2009_Bagheri_Semiconductor_laser_phase_noise_cancellation}%
  \BibitemOpen
  \bibfield  {author} {\bibinfo {author} {\bibfnamefont {M.}~\bibnamefont {Bagheri}}, \bibinfo {author} {\bibfnamefont {F.}~\bibnamefont {Aflatouni}}, \bibinfo {author} {\bibfnamefont {A.}~\bibnamefont {Imani}}, \bibinfo {author} {\bibfnamefont {A.}~\bibnamefont {Goel}},\ and\ \bibinfo {author} {\bibfnamefont {H.}~\bibnamefont {Hashemi}},\ }\href {https://doi.org/10.1364/ol.34.002979} {\bibfield  {journal} {\bibinfo  {journal} {Optics Letters}\ }\textbf {\bibinfo {volume} {34}},\ \bibinfo {pages} {2979} (\bibinfo {year} {2009})}\BibitemShut {NoStop}%
\bibitem [{\citenamefont {Aflatouni}\ and\ \citenamefont {Hashemi}(2012)}]{2012_Aflatouni_Wideband_tunable_laser_phase_noise_reduction}%
  \BibitemOpen
  \bibfield  {author} {\bibinfo {author} {\bibfnamefont {F.}~\bibnamefont {Aflatouni}}\ and\ \bibinfo {author} {\bibfnamefont {H.}~\bibnamefont {Hashemi}},\ }\href {https://doi.org/10.1364/ol.37.000196} {\bibfield  {journal} {\bibinfo  {journal} {Optics Letters}\ }\textbf {\bibinfo {volume} {37}},\ \bibinfo {pages} {196} (\bibinfo {year} {2012})}\BibitemShut {NoStop}%
\bibitem [{\citenamefont {Okamoto}\ \emph {et~al.}(2016)\citenamefont {Okamoto}, \citenamefont {Toge},\ and\ \citenamefont {Manabe}}]{2016_Okamoto_Toge_Manabe_Robust_Laser_Phase_Noise_Reduction}%
  \BibitemOpen
  \bibfield  {author} {\bibinfo {author} {\bibfnamefont {T.}~\bibnamefont {Okamoto}}, \bibinfo {author} {\bibfnamefont {K.}~\bibnamefont {Toge}},\ and\ \bibinfo {author} {\bibfnamefont {T.}~\bibnamefont {Manabe}},\ }\href {https://doi.org/10.1109/jlt.2016.2548359} {\bibfield  {journal} {\bibinfo  {journal} {Journal of Lightwave Technology}\ }\textbf {\bibinfo {volume} {34}},\ \bibinfo {pages} {3908–3913} (\bibinfo {year} {2016})}\BibitemShut {NoStop}%
\bibitem [{\citenamefont {Chen}\ \emph {et~al.}(2019)\citenamefont {Chen}, \citenamefont {Liu},\ and\ \citenamefont {He}}]{2019_HeZuyuan_AOM}%
  \BibitemOpen
  \bibfield  {author} {\bibinfo {author} {\bibfnamefont {J.}~\bibnamefont {Chen}}, \bibinfo {author} {\bibfnamefont {Q.}~\bibnamefont {Liu}},\ and\ \bibinfo {author} {\bibfnamefont {Z.}~\bibnamefont {He}},\ }\href {https://doi.org/10.1109/JLT.2019.2915637} {\bibfield  {journal} {\bibinfo  {journal} {Journal of Lightwave Technology}\ }\textbf {\bibinfo {volume} {37}},\ \bibinfo {pages} {4657} (\bibinfo {year} {2019})}\BibitemShut {NoStop}%
\bibitem [{\citenamefont {Lintz}\ \emph {et~al.}(2017)\citenamefont {Lintz}, \citenamefont {Phung}, \citenamefont {Coulon}, \citenamefont {Faure},\ and\ \citenamefont {L\'ev\'eque}}]{2017_Lintz_Note_MZI_FFFB}%
  \BibitemOpen
  \bibfield  {author} {\bibinfo {author} {\bibfnamefont {M.}~\bibnamefont {Lintz}}, \bibinfo {author} {\bibfnamefont {D.~H.}\ \bibnamefont {Phung}}, \bibinfo {author} {\bibfnamefont {J.-P.}\ \bibnamefont {Coulon}}, \bibinfo {author} {\bibfnamefont {B.}~\bibnamefont {Faure}},\ and\ \bibinfo {author} {\bibfnamefont {T.}~\bibnamefont {L\'ev\'eque}},\ }\href {https://doi.org/10.1063/1.4975400} {\bibfield  {journal} {\bibinfo  {journal} {Review of Scientific Instruments}\ }\textbf {\bibinfo {volume} {88}},\ \bibinfo {pages} {026102} (\bibinfo {year} {2017})}\BibitemShut {NoStop}%
\bibitem [{\citenamefont {Denecker}\ \emph {et~al.}(2025)\citenamefont {Denecker}, \citenamefont {Chew}, \citenamefont {Guillemant}, \citenamefont {Watanabe}, \citenamefont {Tomita}, \citenamefont {Ohmori},\ and\ \citenamefont {de~L\'es\'eleuc}}]{2025_Leseleuc_FF}%
  \BibitemOpen
  \bibfield  {author} {\bibinfo {author} {\bibfnamefont {T.}~\bibnamefont {Denecker}}, \bibinfo {author} {\bibfnamefont {Y.~T.}\ \bibnamefont {Chew}}, \bibinfo {author} {\bibfnamefont {O.}~\bibnamefont {Guillemant}}, \bibinfo {author} {\bibfnamefont {G.}~\bibnamefont {Watanabe}}, \bibinfo {author} {\bibfnamefont {T.}~\bibnamefont {Tomita}}, \bibinfo {author} {\bibfnamefont {K.}~\bibnamefont {Ohmori}},\ and\ \bibinfo {author} {\bibfnamefont {S.}~\bibnamefont {de~L\'es\'eleuc}},\ }\href {https://doi.org/10.1103/PhysRevA.111.042614} {\bibfield  {journal} {\bibinfo  {journal} {Phys. Rev. A}\ }\textbf {\bibinfo {volume} {111}},\ \bibinfo {pages} {042614} (\bibinfo {year} {2025})}\BibitemShut {NoStop}%
\bibitem [{\citenamefont {Chao}\ \emph {et~al.}(2024)\citenamefont {Chao}, \citenamefont {Hua}, \citenamefont {Liang}, \citenamefont {Yue}, \citenamefont {You},\ and\ \citenamefont {Tey}}]{2024_Chao_PDHFeedForward}%
  \BibitemOpen
  \bibfield  {author} {\bibinfo {author} {\bibfnamefont {Y.-X.}\ \bibnamefont {Chao}}, \bibinfo {author} {\bibfnamefont {Z.-X.}\ \bibnamefont {Hua}}, \bibinfo {author} {\bibfnamefont {X.-H.}\ \bibnamefont {Liang}}, \bibinfo {author} {\bibfnamefont {Z.-P.}\ \bibnamefont {Yue}}, \bibinfo {author} {\bibfnamefont {L.}~\bibnamefont {You}},\ and\ \bibinfo {author} {\bibfnamefont {M.~K.}\ \bibnamefont {Tey}},\ }\href {https://doi.org/10.1364/OPTICA.516838} {\bibfield  {journal} {\bibinfo  {journal} {Optica}\ }\textbf {\bibinfo {volume} {11}},\ \bibinfo {pages} {945} (\bibinfo {year} {2024})}\BibitemShut {NoStop}%
\bibitem [{\citenamefont {Chao}\ \emph {et~al.}(2025)\citenamefont {Chao}, \citenamefont {Hua}, \citenamefont {Liang}, \citenamefont {Yue}, \citenamefont {Jia}, \citenamefont {You},\ and\ \citenamefont {Tey}}]{2025_Chao_Robust}%
  \BibitemOpen
  \bibfield  {author} {\bibinfo {author} {\bibfnamefont {Y.-X.}\ \bibnamefont {Chao}}, \bibinfo {author} {\bibfnamefont {Z.-X.}\ \bibnamefont {Hua}}, \bibinfo {author} {\bibfnamefont {X.-H.}\ \bibnamefont {Liang}}, \bibinfo {author} {\bibfnamefont {Z.-P.}\ \bibnamefont {Yue}}, \bibinfo {author} {\bibfnamefont {C.}~\bibnamefont {Jia}}, \bibinfo {author} {\bibfnamefont {L.}~\bibnamefont {You}},\ and\ \bibinfo {author} {\bibfnamefont {M.~K.}\ \bibnamefont {Tey}},\ }\href {https://doi.org/10.1103/PhysRevApplied.23.L011005} {\bibfield  {journal} {\bibinfo  {journal} {Phys. Rev. Appl.}\ }\textbf {\bibinfo {volume} {23}},\ \bibinfo {pages} {L011005} (\bibinfo {year} {2025})}\BibitemShut {NoStop}%
\bibitem [{\citenamefont {Hua}\ \emph {et~al.}(2025)\citenamefont {Hua}, \citenamefont {Chao}, \citenamefont {Jia}, \citenamefont {Liang}, \citenamefont {Yue},\ and\ \citenamefont {Tey}}]{2025_Hua_Feedforward_FDL}%
  \BibitemOpen
  \bibfield  {author} {\bibinfo {author} {\bibfnamefont {Z.-X.}\ \bibnamefont {Hua}}, \bibinfo {author} {\bibfnamefont {Y.-X.}\ \bibnamefont {Chao}}, \bibinfo {author} {\bibfnamefont {C.}~\bibnamefont {Jia}}, \bibinfo {author} {\bibfnamefont {X.-H.}\ \bibnamefont {Liang}}, \bibinfo {author} {\bibfnamefont {Z.-P.}\ \bibnamefont {Yue}},\ and\ \bibinfo {author} {\bibfnamefont {M.~K.}\ \bibnamefont {Tey}},\ }\href {https://doi.org/10.1364/OE.555801} {\bibfield  {journal} {\bibinfo  {journal} {Opt. Express}\ }\textbf {\bibinfo {volume} {33}},\ \bibinfo {pages} {32518} (\bibinfo {year} {2025})}\BibitemShut {NoStop}%
\bibitem [{\citenamefont {Koke}\ \emph {et~al.}(2010)\citenamefont {Koke}, \citenamefont {Grebing}, \citenamefont {Frei}, \citenamefont {Anderson}, \citenamefont {Assion},\ and\ \citenamefont {Steinmeyer}}]{2010_Koke_NatPhys_AOM_FF_Comb}%
  \BibitemOpen
  \bibfield  {author} {\bibinfo {author} {\bibfnamefont {S.}~\bibnamefont {Koke}}, \bibinfo {author} {\bibfnamefont {C.}~\bibnamefont {Grebing}}, \bibinfo {author} {\bibfnamefont {H.}~\bibnamefont {Frei}}, \bibinfo {author} {\bibfnamefont {A.}~\bibnamefont {Anderson}}, \bibinfo {author} {\bibfnamefont {A.}~\bibnamefont {Assion}},\ and\ \bibinfo {author} {\bibfnamefont {G.}~\bibnamefont {Steinmeyer}},\ }\href {https://doi.org/10.1038/nphoton.2010.91} {\bibfield  {journal} {\bibinfo  {journal} {Nature Photonics}\ }\textbf {\bibinfo {volume} {4}},\ \bibinfo {pages} {462} (\bibinfo {year} {2010})}\BibitemShut {NoStop}%
\bibitem [{\citenamefont {Sala}\ \emph {et~al.}(2012)\citenamefont {Sala}, \citenamefont {Gatti}, \citenamefont {Gambetta}, \citenamefont {Coluccelli}, \citenamefont {Galzerano}, \citenamefont {Laporta},\ and\ \citenamefont {Marangoni}}]{2012_Sala_Wide-bandwidth_phase_lock}%
  \BibitemOpen
  \bibfield  {author} {\bibinfo {author} {\bibfnamefont {T.}~\bibnamefont {Sala}}, \bibinfo {author} {\bibfnamefont {D.}~\bibnamefont {Gatti}}, \bibinfo {author} {\bibfnamefont {A.}~\bibnamefont {Gambetta}}, \bibinfo {author} {\bibfnamefont {N.}~\bibnamefont {Coluccelli}}, \bibinfo {author} {\bibfnamefont {G.}~\bibnamefont {Galzerano}}, \bibinfo {author} {\bibfnamefont {P.}~\bibnamefont {Laporta}},\ and\ \bibinfo {author} {\bibfnamefont {M.}~\bibnamefont {Marangoni}},\ }\href {https://doi.org/10.1364/OL.37.002592} {\bibfield  {journal} {\bibinfo  {journal} {Opt. Lett.}\ }\textbf {\bibinfo {volume} {37}},\ \bibinfo {pages} {2592} (\bibinfo {year} {2012})}\BibitemShut {NoStop}%
\bibitem [{\citenamefont {Burkart}\ \emph {et~al.}(2015)\citenamefont {Burkart}, \citenamefont {Sala}, \citenamefont {Kassi}, \citenamefont {Romanini},\ and\ \citenamefont {Marangoni}}]{2015_Burkart_Optical_phase_cloning}%
  \BibitemOpen
  \bibfield  {author} {\bibinfo {author} {\bibfnamefont {J.}~\bibnamefont {Burkart}}, \bibinfo {author} {\bibfnamefont {T.}~\bibnamefont {Sala}}, \bibinfo {author} {\bibfnamefont {S.}~\bibnamefont {Kassi}}, \bibinfo {author} {\bibfnamefont {D.}~\bibnamefont {Romanini}},\ and\ \bibinfo {author} {\bibfnamefont {M.}~\bibnamefont {Marangoni}},\ }\href {https://doi.org/10.1364/OL.40.000816} {\bibfield  {journal} {\bibinfo  {journal} {Opt. Lett.}\ }\textbf {\bibinfo {volume} {40}},\ \bibinfo {pages} {816} (\bibinfo {year} {2015})}\BibitemShut {NoStop}%
\bibitem [{\citenamefont {Watts}\ \emph {et~al.}(2016)\citenamefont {Watts}, \citenamefont {Murdoch},\ and\ \citenamefont {Barry}}]{2016_Watts_Phase_Noise_Reduction_of_an_Optical_Frequency_Comb_beat_MZM}%
  \BibitemOpen
  \bibfield  {author} {\bibinfo {author} {\bibfnamefont {R.~T.}\ \bibnamefont {Watts}}, \bibinfo {author} {\bibfnamefont {S.~G.}\ \bibnamefont {Murdoch}},\ and\ \bibinfo {author} {\bibfnamefont {L.~P.}\ \bibnamefont {Barry}},\ }\href {https://doi.org/10.1109/jphot.2016.2515518} {\bibfield  {journal} {\bibinfo  {journal} {IEEE Photonics Journal}\ }\textbf {\bibinfo {volume} {8}},\ \bibinfo {pages} {1–7} (\bibinfo {year} {2016})}\BibitemShut {NoStop}%
\bibitem [{\citenamefont {Sahni}\ \emph {et~al.}(2017)\citenamefont {Sahni}, \citenamefont {Trebaol}, \citenamefont {Bramerie}, \citenamefont {Joindot}, \citenamefont {D\'{u}ill}, \citenamefont {Murdoch}, \citenamefont {Barry},\ and\ \citenamefont {Besnard}}]{2017_Besnard_beat_MZM}%
  \BibitemOpen
  \bibfield  {author} {\bibinfo {author} {\bibfnamefont {M.~O.}\ \bibnamefont {Sahni}}, \bibinfo {author} {\bibfnamefont {S.}~\bibnamefont {Trebaol}}, \bibinfo {author} {\bibfnamefont {L.}~\bibnamefont {Bramerie}}, \bibinfo {author} {\bibfnamefont {M.}~\bibnamefont {Joindot}}, \bibinfo {author} {\bibfnamefont {S.~P.~O.}\ \bibnamefont {D\'{u}ill}}, \bibinfo {author} {\bibfnamefont {S.~G.}\ \bibnamefont {Murdoch}}, \bibinfo {author} {\bibfnamefont {L.~P.}\ \bibnamefont {Barry}},\ and\ \bibinfo {author} {\bibfnamefont {P.}~\bibnamefont {Besnard}},\ }\href {https://doi.org/10.1364/OL.42.004000} {\bibfield  {journal} {\bibinfo  {journal} {Opt. Lett.}\ }\textbf {\bibinfo {volume} {42}},\ \bibinfo {pages} {4000} (\bibinfo {year} {2017})}\BibitemShut {NoStop}%
\bibitem [{\citenamefont {Yang}\ \emph {et~al.}(2020)\citenamefont {Yang}, \citenamefont {Lv}, \citenamefont {Luo}, \citenamefont {Hu}, \citenamefont {Yang}, \citenamefont {Fu},\ and\ \citenamefont {Tan}}]{2020-Tan-Sensor-beat-FFFB-AOM}%
  \BibitemOpen
  \bibfield  {author} {\bibinfo {author} {\bibfnamefont {R.}~\bibnamefont {Yang}}, \bibinfo {author} {\bibfnamefont {H.}~\bibnamefont {Lv}}, \bibinfo {author} {\bibfnamefont {J.}~\bibnamefont {Luo}}, \bibinfo {author} {\bibfnamefont {P.}~\bibnamefont {Hu}}, \bibinfo {author} {\bibfnamefont {H.}~\bibnamefont {Yang}}, \bibinfo {author} {\bibfnamefont {H.}~\bibnamefont {Fu}},\ and\ \bibinfo {author} {\bibfnamefont {J.}~\bibnamefont {Tan}},\ }\href {https://doi.org/10.3390/s20051248} {\bibfield  {journal} {\bibinfo  {journal} {Sensors}\ }\textbf {\bibinfo {volume} {20}},\ \bibinfo {pages} {1248} (\bibinfo {year} {2020})}\BibitemShut {NoStop}%
\bibitem [{\citenamefont {Li}\ \emph {et~al.}(2022)\citenamefont {Li}, \citenamefont {Huie}, \citenamefont {Chen}, \citenamefont {DeMarco},\ and\ \citenamefont {Covey}}]{2022_Li_Active_Cancellation_of_Servo-Induced_Noise}%
  \BibitemOpen
  \bibfield  {author} {\bibinfo {author} {\bibfnamefont {L.}~\bibnamefont {Li}}, \bibinfo {author} {\bibfnamefont {W.}~\bibnamefont {Huie}}, \bibinfo {author} {\bibfnamefont {N.}~\bibnamefont {Chen}}, \bibinfo {author} {\bibfnamefont {B.}~\bibnamefont {DeMarco}},\ and\ \bibinfo {author} {\bibfnamefont {J.~P.}\ \bibnamefont {Covey}},\ }\href {https://doi.org/10.1103/PhysRevApplied.18.064005} {\bibfield  {journal} {\bibinfo  {journal} {Phys. Rev. Appl.}\ }\textbf {\bibinfo {volume} {18}},\ \bibinfo {pages} {064005} (\bibinfo {year} {2022})}\BibitemShut {NoStop}%
\bibitem [{\citenamefont {Woo}\ and\ \citenamefont {Kim}(2024)}]{2024_Beat_IQmodulator}%
  \BibitemOpen
  \bibfield  {author} {\bibinfo {author} {\bibfnamefont {K.}~\bibnamefont {Woo}}\ and\ \bibinfo {author} {\bibfnamefont {H.}~\bibnamefont {Kim}},\ }\href {https://doi.org/10.1109/LPT.2024.3379496} {\bibfield  {journal} {\bibinfo  {journal} {IEEE Photonics Technology Letters}\ }\textbf {\bibinfo {volume} {36}},\ \bibinfo {pages} {563} (\bibinfo {year} {2024})}\BibitemShut {NoStop}%
\bibitem [{\citenamefont {Maddox}\ \emph {et~al.}(2024)\citenamefont {Maddox}, \citenamefont {Mortlock}, \citenamefont {Hepworth}, \citenamefont {Raghuram}, \citenamefont {Gregory}, \citenamefont {Guttridge},\ and\ \citenamefont {Cornish}}]{2025_Maddox_Enhanced_Quantum_State_Transfer}%
  \BibitemOpen
  \bibfield  {author} {\bibinfo {author} {\bibfnamefont {B.~P.}\ \bibnamefont {Maddox}}, \bibinfo {author} {\bibfnamefont {J.~M.}\ \bibnamefont {Mortlock}}, \bibinfo {author} {\bibfnamefont {T.~R.}\ \bibnamefont {Hepworth}}, \bibinfo {author} {\bibfnamefont {A.~P.}\ \bibnamefont {Raghuram}}, \bibinfo {author} {\bibfnamefont {P.~D.}\ \bibnamefont {Gregory}}, \bibinfo {author} {\bibfnamefont {A.}~\bibnamefont {Guttridge}},\ and\ \bibinfo {author} {\bibfnamefont {S.~L.}\ \bibnamefont {Cornish}},\ }\href {https://doi.org/10.1103/PhysRevLett.133.253202} {\bibfield  {journal} {\bibinfo  {journal} {Phys. Rev. Lett.}\ }\textbf {\bibinfo {volume} {133}},\ \bibinfo {pages} {253202} (\bibinfo {year} {2024})}\BibitemShut {NoStop}%
\bibitem [{sta()}]{stabilization_note}%
  \BibitemOpen
  \href@noop {} {}\bibinfo {note} {This stabilization is included after we discovered drift of the relative phase between LO2 and the beat signal even when OPL is active. The exact reason is unclear but is potentially due to close to mode-hop operations in our semiconductor slave laser. Its inclusion nevertheless allows for auto phase adjustment when ramping $\Delta \omega$.}\BibitemShut {Stop}%
\end{thebibliography}

\end{document}